\documentclass[prb,preprint]{revtex4-1} 
\usepackage{amsmath}  % needed for \tfrac, \bmatrix, etc.
\usepackage{amsfonts} % needed for bold Greek, Fraktur, and blackboard bold
\usepackage{graphicx} % needed for figures
\newcommand{\unit}[1]{\ensuremath{\, \mathrm{#1}}}

\begin{document}

% Be sure to use the \title, \author, \affiliation, and \abstract macros
% to format your title page.  Don't use lower-level macros to  manually
% adjust the fonts and centering.

\title{Calculating the Potato Radius of Asteroids using the Height of Mt. Everest}
% In a long title you can use \\ to force a line break at a certain location.

\author{M. E. Caplan}
\email{mecaplan@indiana.edu} % optional
%\altaffiliation[permanent address: ]{2401 N. Milo B. Sampson Lane, Bloomington, IN 47401} % optional second address
% If there were a second author at the same address, we would put another 
% \author{} statement here.  Don't combine multiple authors in a single
% \author statement.
\affiliation{Center for the Exploration of Energy and Matter, Indiana University, Bloomington, IN 47408}
% Please provide a full mailing address here.

%\author{David P. Jackson}
%\email{ajp@dickinson.edu}
%\affiliation{Department of Physics, Dickinson College, Carlisle, PA 17013}

% See the REVTeX documentation for more examples of author and affiliation lists.

\date{\today}

\begin{abstract}
At approximate radii of 200-300 km, asteroids transition from oblong `potato' shapes to spheres. This limit is known as the Potato Radius, and has been proposed as a classification for separating asteroids from dwarf planets. The Potato Radius can be calculated from first principles based on the elastic properties and gravity of the asteroid. Similarly, the tallest mountain that a planet can support is also known to be based on the elastic properties and gravity. In this work, a simple novel method of calculating the Potato Radius is presented using what is known about the maximum height of mountains and Newtonian gravity for a spherical body. This method does not assume any knowledge beyond high school level mechanics, and may be appropriate for students interested in applications of physics to astronomy. 
\end{abstract}
% AJP requires an abstract for all regular article submissions.
% Abstracts are optional for submissions to the "Notes and Discussions" section.

\maketitle % title page is now complete

\section{Introduction} % Section titles are automatically converted to all-caps.
% Section numbering is automatic.
%

%The NEAR spacecraft landed on the potato-shaped asteroid Eros in 2000, while Dawn has orbited the minor planets Vesta and Ceres, and New Horizons flew by Pluto and Charon. 

\begin{figure}[ht!]
\includegraphics[width=125mm]{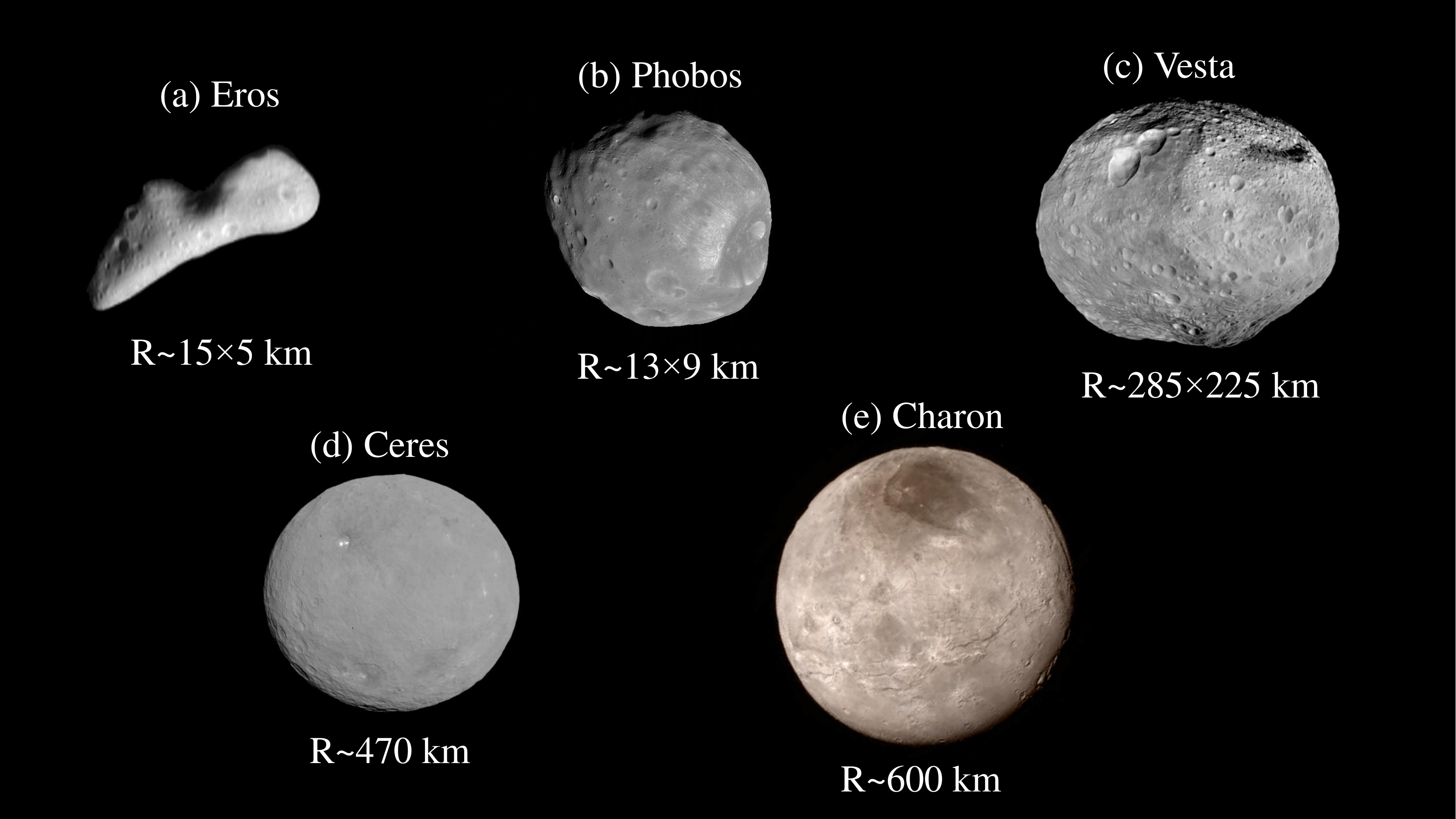}
\caption{\label{fig:satelliteimages} Five solar system minor bodies imaged by visiting probes. Clockwise from top left: (a) Eros as seen by the NEAR mission, (b) Phobos as seen by the MRO, (c) Vesta and (d) Ceres as seen by the Dawn mission, and (e) Charon seen by New Horizons. Approximate radii from Ref. \cite{NASA-AsteroidFactSheet}.}
\end{figure}

Spacecraft are currently exploring asteroids and dwarf planets, such as the Near Earth Asteroid Rendezvous mission (NEAR) landing on Eros,\cite{NASA-NEAR} the Dawn mission orbiting Ceres and Vesta,\cite{NASA-DAWN} and the New Horizons flyby of Pluto and Charon.\cite{NASA-NH} Additionally, the Mars Reconnaissance Orbiter (MRO) has observed the Martian moons Phobos and Deimos.\cite{NASA-MRO} These missions observe a remarkable variety of shapes for these bodies, shown in Fig. \ref{fig:satelliteimages}. Smaller asteroids have irregular shapes while dwarf planets (large asteroids) are nearly spherical. This follows from some simple physics. 

A well known result of Newtonian gravity is that a material with uniform density has the minimum gravitational potential energy when shaped into a sphere. 
%spherical distribution of mass has the lowest potential energy for a constant density material. 
If a sufficiently deep hole is dug in a spherical body, material will fall in from the edges. If matter is stacked into a sufficiently high mountain it will eventually fall down under its own weight. 

For a sufficiently large surface deviation from spherical the gravitational force will be able to overcome the material's yield strength and deform it back to some maximum allowable deviation, determined by the strength of the force of gravity and the material's elastic properties. Thus, planets are spherical with small surface deviations. 

%Simply put, the sphere is the only shape without holes to fill or mountains to crust. 

In reality, all large isolated bodies that have been observed are found to be nearly spherical, or at least oblate spheroidal with an equatorial bulge due to rotation. However, many asteroids and moons in the solar system with radii less than 200-300 km are known to have oblong and asymmetric geometries.\cite{Lineweaver} This is because these asteroids and moons do not have sufficient gravity to overcome their intrinsic rigidity, and can thus maintain their nonspherical shapes. This radius range associated with the transition between oblong and spherical geometries has been dubbed `The Potato Radius.' 
%The limit in radius associated with the observed transition between oblong and spherical geometries at radii of approximately 200-300 km has been dubbed `The Potato Radius.' 

Additionally, the highest possible mountain that a planet can support has been studied from first principles, where it has been found that the maximum height of a mountain is dependent on the surface gravity and elastic properties of the mountain.\cite{Scheuer} 

As a simple illustrative argument, the maximum height of a mountain on a body of uniform density is limited by the yield strength $S$ of the mountain. The pressure at the base of the mountain is $\sim\rho g h $, where $\rho$ is the density, $g$ the surface gravity, and $h$ the height of the mountain. If the height is such that $\rho g h \gtrsim S$ the base will break, causing the mountain to crumble back to the maximum allowable height.\cite{cole2013planetary} Rearranging terms, we can find that $h g \sim S/\rho$, implying that the product of the tallest possible mountain and the surface gravity is constant. If a pair of bodies are made of the same material %$S$ and $\rho$ are equal, so 
an approximate relation arises which we call the Height-Gravity relation: 

\begin{equation}
h_1 g_1 = h_2 g_2 = C
\label{eqn:hg}
\end{equation}

\noindent Where bodies 1 and 2 have respective surface gravities $g_1$ and $g_2$, and maximum mountain heights of $h_1$ and $h_2$, and $C\sim S/\rho$ is taken to be a constant dependent on composition which we call the Rock Constant in this work. \footnote{A similar constant could be derived for ice which could be used, for example, to determine the maximum height of cryovolcanoes on various moons. Similarly, a constant could be derived using the shear modulus of nuclear matter to find the maximum height of mountains on neutron stars.} For two bodies with similar compositions, Eq. \ref{eqn:hg} implies larger planets will have a smaller tallest possible mountain while smaller planets can support larger mountains.  

This relation is obeyed quite well in the inner solar system. For example, the surface gravity of Mars is $\approx (2/5) g_{Earth}$ and the height of Olympus Mons (the tallest mountain on Mars) is nearly $(5/2) h_{Everest}$. For rocky bodies in the inner solar system, the heights of the tallest mountains are given in Tab. \ref{tab:mountain_heights}.

\begingroup
\squeezetable
\begin{table}[h]
%\centering
\caption{\label{tab:mountain_heights} List of the tallest mountains on each planet, taken to be the highest point above mean surface elevation. The Rock Constant of the Height-Gravity relation is calculated from the height of the tallest mountain and surface gravity of each planet.}
\label{my-label}
\begin{ruledtabular}
\begin{tabular}{ccccc}
Planet  & Tallest Mountain & Height above mean     & Surface gravity\cite{NASA-PlanetaryFactSheet} ($\frac{m}{s^2}$) & Rock Constant ($\frac{m^2}{s^2}$) \\
        &                  &  planetary radius (m)  &            &       \\
\hline
Mercury & Caloris Montes &   $>$3000 \cite{Oberst2010230}       &    3.7      &  11100    \\
Venus   & Maxwell Montes &   10670   \cite{1992JGR....9716353K}       &    8.9      &  94963    \\
Earth   & Mt Everest     &   8850           &    9.8      &  86730     \\
Mars    & Olympus Mons   &   21900   \cite{2004JGRE..109.3003P}       &    3.7      &  81030    
\end{tabular}
\end{ruledtabular}
\end{table}
\endgroup

The product $hg$ gives a value for the Rock Constant that is constant to $\sim 10\%$ for Venus, Earth, and Mars. The exception is Mercury, whose tallest mountain falls well below the limit. Recall that the Height-Gravity relation provides an upper-bound - the tallest mountain on Mercury is simply not the tallest possible. Similarly, the tallest mountain on the Moon and other rocky bodies in the solar system are found to be well below the limit implied by Height-Gravity relation. This could be because Mercury and the Moon are not geologically active. Caloris Montes on Mercury is the rim of an impact crater,\cite{Oberst2010230}  while Maxwell Montes, Mt. Everest, and Olympus Mons were all produced by volcanism or tectonic activity.\cite{1992JGR....9716353K,2004JGRE..109.3003P}

Because the constant in the Height-Gravity relation and the Potato Radius are both dependent on material specific constants, the Height-Gravity relation can be used to derive the Potato Radius. This derivation could be of interest to introductory physics students with an interest in applications of physics to astronomy, particularly because it does not assume an understanding of advanced calculus or material science (e.g. Young's Modulus, Stress, Yield Strength, etc) that are required to understand previous derivations of the Potato Radius. 

%Because the Height-Gravity relation implies that the tallest possible mountain is dependent only on surface gravity and material specific constants we can use the Height-Gravity relation to derive the Potato Radius. 

\section{Calculating the Potato Radius}

Recall that the Potato Radius is the radius where there is a transition from oblong asteroids to spherical dwarf planets. If we consider an asteroid as a small sphere with large surface deviations we can apply the Height-Gravity relation to find the tallest possible mountain. As the radius of this sphere increases the surface gravity increases, and thus the maximum mountain height eventually decreases below the radius, and the asteroid becomes nearly spherical. The radius where the maximum height of a mountain is equal to the radius of the asteroid should therefore be the Potato Radius. 

\begin{figure}[ht!]
\includegraphics[width=50mm]{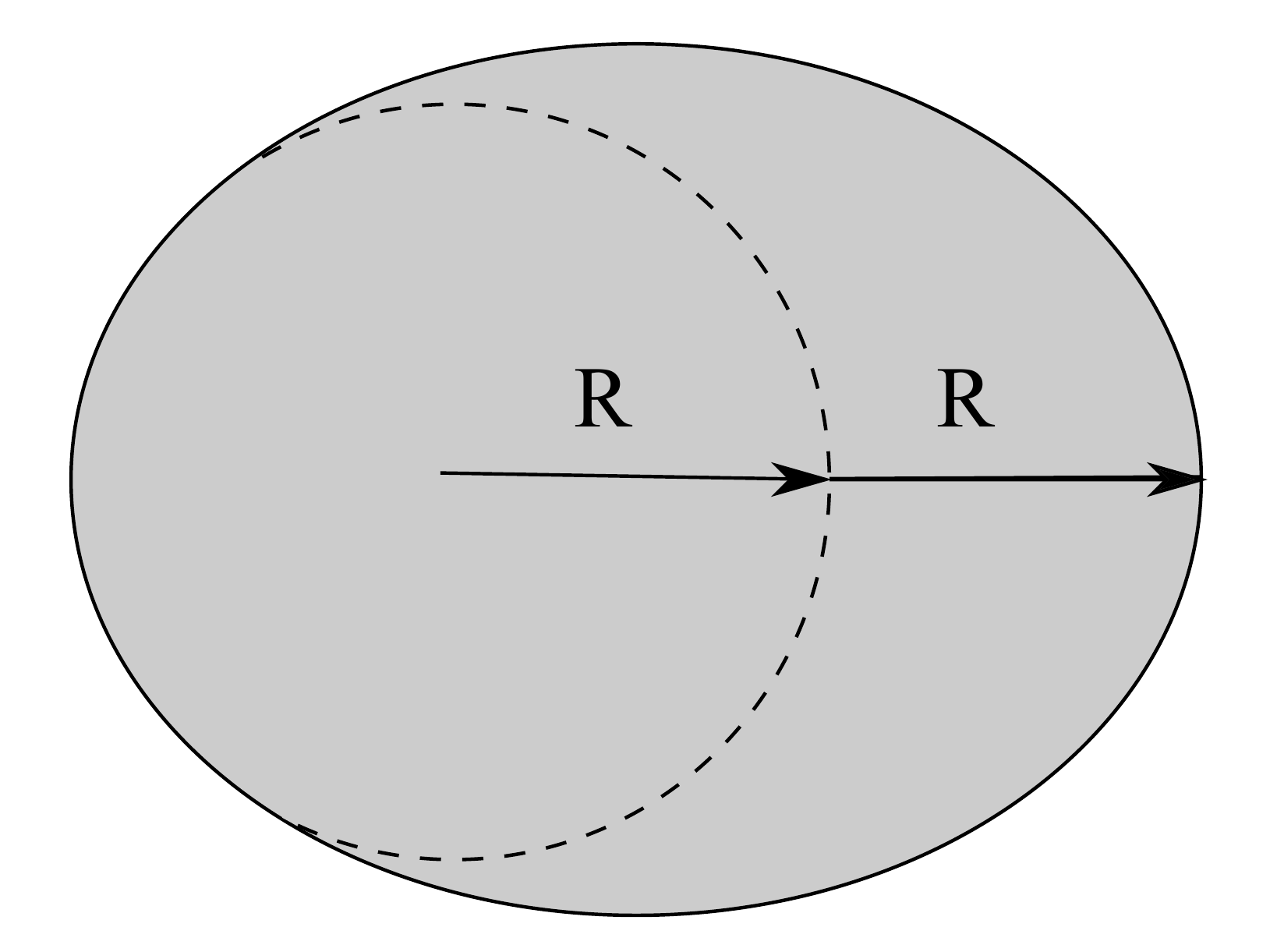}
\caption{\label{fig:potato_asteroid} Oblong asteroid approximated as a sphere of radius $R$ with a hemisphere-spanning mountain of height $R$.}
\end{figure}

Consider an ellipsoidal asteroid with a semi-minor axis $R$ and a semi-major axis $1.5R$. This asteroid could be approximated by a sphere of radius R with a mountain of height $R$ covering one hemisphere, as shown in Fig. \ref{fig:potato_asteroid}.

\noindent The Height-Gravity relation can be applied

\begin{equation}
h_{asteroid} \, g_{asteroid} = C
\end{equation}

\noindent and the Rock Constant $C$ can be taken to be the product of the height of Everest with earth surface gravity - this value was nearly the mean of the possible values calculated in Tab. \ref{tab:mountain_heights}. The height $h_{asteroid}$ of the mountain is already taken to be the radius $R$, and the surface gravity of the sphere can be found by Newton's Law of Gravitation

\begin{equation}
g = \frac{G M}{R^2}
\end{equation}

\noindent and by taking the asteroid to have a constant density $\rho$ this can be expressed purely in terms of the radius, 

\begin{equation}
g = \frac{G (\frac{4}{3}\pi\rho R^3)}{R^2} = \frac{4}{3}\ G \rho R.
\end{equation}

At this point, the Height-Gravity relation can be applied:

\begin{eqnarray}
h_{asteroid} g_{asteroid} = h_{Earth} g_{Earth} \nonumber \\
R (\frac{4}{3} \pi G \rho R) = h_{Earth} g_{Earth}  \\
R = \sqrt{\frac{3 h_{Earth} g_{Earth}}{4\pi G \rho}} . \nonumber
\end{eqnarray}

\noindent Using $h_{Earth} = 8850 \unit{m}$, $g_{Earth}=9.8 \unit{m/s}^2$ , and $\rho = 5.5 \unit{g/cm}^3$ ($\approx \rho_{Earth}$), this equation gives a value of $R=238 \unit{km}$. This is within the range of 200-300 km calculated directly by Lineweaver and Norman, and in agreement with observation of the oblong shapes of Eros, Phobos, and Vesta are irregular while larger bodies like Ceres and Charon are nearly spherical. 

Slightly different values of the Potato Radius can be obtained by using a more realistic density for the asteroid (there doesn't seem to be any reason it should be the same density as the earth), though even the most well constrained asteroid densities still vary between $\sim 2-10 \unit{g/cm}^3$.\cite{carry2012density} 
Furthermore, %the assumption of a constant surface gravity may not be well founded.
on a large planet $R \gg h$ so surface gravity is approximately constant over the span of the mountain. In contrast, when $h\sim R$ the surface gravity may vary considerably over the mountain, thus requiring a more rigorous treatment.  
Lastly, the choice of Everest and earth gravity for the Rock Constant applies for rocky bodies of earth like composition, while objects of different composition will have a different constant.

\section{Conclusion}

Objects larger than the Potato Radius must be nearly spherical, while objects smaller objects can be asymmetric.
In this work, a novel method was presented for calculating the Potato Radius using the maximum height of mountains on planets. Our result for the Potato Radius $R\approx 240 \unit{km}$ agrees well with spacecraft observation. This method assumes no knowledge beyond introductory mechanics, and may be of interest to students and teachers interested in practical applications of physics to astronomy. Unlike previous methods for calculating the Potato Radius and the maximum height of mountains, this method does not require extensive knowledge of calculus or materials physics, such as the Young's modulus, stress, and yield strength. 

\section{Acknowledgments}

The author would like to thank C. J. Horowitz for useful discussion. 

\bibliography{potato_radius}

\begin{thebibliography}{14}%
\makeatletter
\providecommand \@ifxundefined [1]{%
 \@ifx{#1\undefined}
}%
\providecommand \@ifnum [1]{%
 \ifnum #1\expandafter \@firstoftwo
 \else \expandafter \@secondoftwo
 \fi
}%
\providecommand \@ifx [1]{%
 \ifx #1\expandafter \@firstoftwo
 \else \expandafter \@secondoftwo
 \fi
}%
\providecommand \natexlab [1]{#1}%
\providecommand \enquote  [1]{``#1''}%
\providecommand \bibnamefont  [1]{#1}%
\providecommand \bibfnamefont [1]{#1}%
\providecommand \citenamefont [1]{#1}%
\providecommand \href@noop [0]{\@secondoftwo}%
\providecommand \href [0]{\begingroup \@sanitize@url \@href}%
\providecommand \@href[1]{\@@startlink{#1}\@@href}%
\providecommand \@@href[1]{\endgroup#1\@@endlink}%
\providecommand \@sanitize@url [0]{\catcode `\\12\catcode `\$12\catcode
  `\&12\catcode `\#12\catcode `\^12\catcode `\_12\catcode `\%12\relax}%
\providecommand \@@startlink[1]{}%
\providecommand \@@endlink[0]{}%
\providecommand \url  [0]{\begingroup\@sanitize@url \@url }%
\providecommand \@url [1]{\endgroup\@href {#1}{\urlprefix }}%
\providecommand \urlprefix  [0]{URL }%
\providecommand \Eprint [0]{\href }%
\providecommand \doibase [0]{http://dx.doi.org/}%
\providecommand \selectlanguage [0]{\@gobble}%
\providecommand \bibinfo  [0]{\@secondoftwo}%
\providecommand \bibfield  [0]{\@secondoftwo}%
\providecommand \translation [1]{[#1]}%
\providecommand \BibitemOpen [0]{}%
\providecommand \bibitemStop [0]{}%
\providecommand \bibitemNoStop [0]{.\EOS\space}%
\providecommand \EOS [0]{\spacefactor3000\relax}%
\providecommand \BibitemShut  [1]{\csname bibitem#1\endcsname}%
\let\auto@bib@innerbib\@empty
%</preamble>
\bibitem [{NAS(2001{\natexlab{a}})}]{NASA-NEAR}%
  \BibitemOpen
  \href@noop {} {\enquote {\bibinfo {title} {{NASA GSFC - NEAR}},}\ }\bibinfo
  {howpublished}
  {\url{http://nssdc.gsfc.nasa.gov/planetary/mission/near/near_eros.html}}
  (\bibinfo {year} {2001}{\natexlab{a}}),\ \bibinfo {note} {accessed:
  2015-11-05}\BibitemShut {NoStop}%
\bibitem [{NAS(2015{\natexlab{a}})}]{NASA-DAWN}%
  \BibitemOpen
  \href@noop {} {\enquote {\bibinfo {title} {{NASA JPL - Dawn Mission}},}\
  }\bibinfo {howpublished} {\url{http://dawn.jpl.nasa.gov/}} (\bibinfo {year}
  {2015}{\natexlab{a}}),\ \bibinfo {note} {accessed: 2015-11-05}\BibitemShut
  {NoStop}%
\bibitem [{NAS(2015{\natexlab{b}})}]{NASA-NH}%
  \BibitemOpen
  \href@noop {} {\enquote {\bibinfo {title} {{Johns Hopkins University Applied
  Physics Laboratory - New Horizons}},}\ }\bibinfo {howpublished}
  {\url{http://pluto.jhuapl.edu/}} (\bibinfo {year} {2015}{\natexlab{b}}),\
  \bibinfo {note} {accessed: 2015-11-05}\BibitemShut {NoStop}%
\bibitem [{NAS(2015{\natexlab{c}})}]{NASA-MRO}%
  \BibitemOpen
  \href@noop {} {\enquote {\bibinfo {title} {{NASA - Mars Reconnaissance
  Orbiter}},}\ }\bibinfo {howpublished}
  {\url{http://www.nasa.gov/mission_pages/MRO/main/index.html}} (\bibinfo
  {year} {2015}{\natexlab{c}}),\ \bibinfo {note} {accessed:
  2015-11-05}\BibitemShut {NoStop}%
\bibitem [{NAS(2001{\natexlab{b}})}]{NASA-AsteroidFactSheet}%
  \BibitemOpen
  \href@noop {} {\enquote {\bibinfo {title} {{NASA GSFC - Asteroid Fact
  Sheet}},}\ }\bibinfo {howpublished}
  {\url{http://nssdc.gsfc.nasa.gov/planetary/factsheet/asteroidfact.html}}
  (\bibinfo {year} {2001}{\natexlab{b}}),\ \bibinfo {note} {accessed:
  2015-11-05}\BibitemShut {NoStop}%
\bibitem [{\citenamefont {{Lineweaver}}\ and\ \citenamefont
  {{Norman}}(2009)}]{Lineweaver}%
  \BibitemOpen
  \bibfield  {author} {\bibinfo {author} {\bibfnamefont {C.~H.}\ \bibnamefont
  {{Lineweaver}}}\ and\ \bibinfo {author} {\bibfnamefont {M.}~\bibnamefont
  {{Norman}}},\ }\href@noop {} {\bibfield  {journal} {\bibinfo  {journal}
  {Australian Space Science Conference Series: Proceedings of the 9th
  Australian Space Science Conference}\ ,\ \bibinfo {pages} {67}} (\bibinfo
  {year} {2009})}\BibitemShut {NoStop}%
\bibitem [{\citenamefont {Scheuer}(1981)}]{Scheuer}%
  \BibitemOpen
  \bibfield  {author} {\bibinfo {author} {\bibfnamefont {P.}~\bibnamefont
  {Scheuer}},\ }\href {\doibase 10.1007/BF02715676} {\bibfield  {journal}
  {\bibinfo  {journal} {Journal of Astrophysics and Astronomy}\ }\textbf
  {\bibinfo {volume} {2}},\ \bibinfo {pages} {165} (\bibinfo {year}
  {1981})}\BibitemShut {NoStop}%
\bibitem [{\citenamefont {Cole}\ and\ \citenamefont
  {Woolfson}(2013)}]{cole2013planetary}%
  \BibitemOpen
  \bibfield  {author} {\bibinfo {author} {\bibfnamefont {G.}~\bibnamefont
  {Cole}}\ and\ \bibinfo {author} {\bibfnamefont {M.}~\bibnamefont
  {Woolfson}},\ }\href {https://books.google.com/books?id=B17OBQAAQBAJ} {\emph
  {\bibinfo {title} {Planetary Science: The Science of Planets around Stars,
  Second Edition}}}\ (\bibinfo  {publisher} {CRC Press},\ \bibinfo {year}
  {2013})\BibitemShut {NoStop}%
\bibitem [{Note1()}]{Note1}%
  \BibitemOpen
  \bibinfo {note} {A similar constant could be derived for ice which could be
  used, for example, to determine the maximum height of cryovolcanoes on
  various moons. Similarly, a constant could be derived using the shear modulus
  of nuclear matter to find the maximum height of mountains on neutron
  stars.}\BibitemShut {Stop}%
\bibitem [{NAS(2015{\natexlab{d}})}]{NASA-PlanetaryFactSheet}%
  \BibitemOpen
  \href@noop {} {\enquote {\bibinfo {title} {{NASA GSFC - Planetary Fact
  Sheet}},}\ }\bibinfo {howpublished}
  {\url{http://nssdc.gsfc.nasa.gov/planetary/factsheet/}} (\bibinfo {year}
  {2015}{\natexlab{d}}),\ \bibinfo {note} {accessed: 2015-11-04}\BibitemShut
  {NoStop}%
\bibitem [{\citenamefont {Oberst}\ \emph {et~al.}(2010)\citenamefont {Oberst},
  \citenamefont {Preusker}, \citenamefont {Phillips}, \citenamefont {Watters},
  \citenamefont {Head}, \citenamefont {Zuber},\ and\ \citenamefont
  {Solomon}}]{Oberst2010230}%
  \BibitemOpen
  \bibfield  {author} {\bibinfo {author} {\bibfnamefont {J.}~\bibnamefont
  {Oberst}}, \bibinfo {author} {\bibfnamefont {F.}~\bibnamefont {Preusker}},
  \bibinfo {author} {\bibfnamefont {R.~J.}\ \bibnamefont {Phillips}}, \bibinfo
  {author} {\bibfnamefont {T.~R.}\ \bibnamefont {Watters}}, \bibinfo {author}
  {\bibfnamefont {J.~W.}\ \bibnamefont {Head}}, \bibinfo {author}
  {\bibfnamefont {M.~T.}\ \bibnamefont {Zuber}}, \ and\ \bibinfo {author}
  {\bibfnamefont {S.~C.}\ \bibnamefont {Solomon}},\ }\href {\doibase
  http://dx.doi.org/10.1016/j.icarus.2010.03.009} {\bibfield  {journal}
  {\bibinfo  {journal} {Icarus}\ }\textbf {\bibinfo {volume} {209}},\ \bibinfo
  {pages} {230 } (\bibinfo {year} {2010})},\ \bibinfo {note} {{Mercury after
  Two MESSENGER Flybys }}\BibitemShut {NoStop}%
\bibitem [{\citenamefont {{Klose}}\ \emph {et~al.}(1992)\citenamefont
  {{Klose}}, \citenamefont {{Wood}},\ and\ \citenamefont
  {{Hashimoto}}}]{1992JGR....9716353K}%
  \BibitemOpen
  \bibfield  {author} {\bibinfo {author} {\bibfnamefont {K.~B.}\ \bibnamefont
  {{Klose}}}, \bibinfo {author} {\bibfnamefont {J.~A.}\ \bibnamefont {{Wood}}},
  \ and\ \bibinfo {author} {\bibfnamefont {A.}~\bibnamefont {{Hashimoto}}},\
  }\href {\doibase 10.1029/92JE01865} {\bibfield  {journal} {\bibinfo
  {journal} {Journal of Geophysical Research}\ }\textbf {\bibinfo {volume}
  {97}},\ \bibinfo {pages} {16353} (\bibinfo {year} {1992})}\BibitemShut
  {NoStop}%
\bibitem [{\citenamefont {{Plescia}}(2004)}]{2004JGRE..109.3003P}%
  \BibitemOpen
  \bibfield  {author} {\bibinfo {author} {\bibfnamefont {J.~B.}\ \bibnamefont
  {{Plescia}}},\ }\href {\doibase 10.1029/2002JE002031} {\bibfield  {journal}
  {\bibinfo  {journal} {Journal of Geophysical Research (Planets)}\ }\textbf
  {\bibinfo {volume} {109}},\ \bibinfo {eid} {E03003} (\bibinfo {year}
  {2004})}\BibitemShut {NoStop}%
\bibitem [{\citenamefont {Carry}(2012)}]{carry2012density}%
  \BibitemOpen
  \bibfield  {author} {\bibinfo {author} {\bibfnamefont {B.}~\bibnamefont
  {Carry}},\ }\href@noop {} {\bibfield  {journal} {\bibinfo  {journal}
  {Planetary and Space Science}\ }\textbf {\bibinfo {volume} {73}},\ \bibinfo
  {pages} {98} (\bibinfo {year} {2012})}\BibitemShut {NoStop}%
\end{thebibliography}%

\end{document}